\documentclass[aps,prd,onecolumn,nofootinbib,superscriptaddress]{revtex4}
\usepackage{float}
\usepackage{graphicx}
\usepackage{amsmath}
\usepackage{amsfonts}
\usepackage{amssymb,ulem}
\usepackage{color}%
\usepackage{dcolumn}
\usepackage{subfigure}
\usepackage{multirow}

\usepackage{MnSymbol,wasysym}
\usepackage{braket,diagbox}
\usepackage{eurosym}
\usepackage{calrsfs}
\usepackage[usenames,dvipsnames,svgnames]{xcolor}

\newcommand{\RNum}[1]{\uppercase\expandafter{\romannumeral #1\relax}}
\usepackage[colorlinks=true,linkcolor=blue,urlcolor=blue,filecolor=black,citecolor=red,
pdfstartview=FitV,pdftitle={},pdfsubject={},pdfkeywords={},pdfpagemode=None,bookmarksopen=true]{hyperref}
\usepackage{float}

\begin{document}
\baselineskip=0.4 cm

\title{Energy Extraction via Magnetic Reconnection from Rotating ModMax Black Holes}
\author{Shi-Hao Wang}
\email{shwangphys@163.com}
\affiliation{School of Physics and Astronomy, China West Normal University, Nanchong 637000, China}

\author{Ke-Jian He}
\email{kjhe94@163.com}
\affiliation{Department of Mechanics, Chongqing Jiaotong University, Chongqing 400000, China}

\author{Guo-Ping Li}
\email{gpliphys@yeah.net}
\affiliation{School of Physics and Astronomy, China West Normal University, Nanchong 637000, China}

\author{Yuan Meng}
\email{mengyuanphy@163.com (Corresponding author)}
\affiliation{School of Physics and Astronomy, China West Normal University, Nanchong 637000, China}

\date{\today}

\begin{abstract}
\baselineskip=0.5 cm

Magnetic reconnection has been widely recognized as an important mechanism for extracting energy from rapidly rotating black holes. We find that magnetic reconnection can also efficiently extract energy from slowly rotating ModMax black holes. In this paper, we investigate the magnetic reconnection process of ModMax black holes in both circular and plunging orbits. First, we analyze the fundamental characteristic quantities of the ModMax black hole, including the event horizon, ergosphere radius, and circular photon orbits. The results indicate that the charge parameter $Q$ and screening factor $\gamma$ exhibit a competing effect on the fundamental characteristic quantities of the ModMax black hole. Furthermore, for the extremal ModMax black hole, a larger $Q$ and a smaller $\gamma$ lower the minimum allowed spin parameter $a$. Subsequently, we analyze the parameter space $(r,a)$ for energy extraction in circular and plunging orbits. A larger $Q$ and a smaller $\gamma$ reduce the region for energy extraction and decrease the minimum spin parameter $a$ required for energy extraction. For circular orbits, the minimum allowed spin parameter is $a\simeq 0.50071$, while for plunging orbits, this threshold further decreases to $a\simeq0.22407$. This proves that energy extraction from lower spin ModMax black holes is theoretically feasible. Finally, we compare the energy extraction region, power, and efficiency between circular and plunging orbits. It is found that these quantities in plunging orbits are always higher than those in circular orbits, indicating that energy extraction from plunging orbits via magnetic reconnection may be more efficient.

\end{abstract}


\maketitle
\tableofcontents

\section{Introduction}

Black holes, as mysterious celestial objects predicted by general relativity, have long attracted extensive attention. The Laser Interferometer Gravitational-wave Observatory (LIGO) has for the first time detected gravitational wave signals from the merger of two black holes, providing empirical evidence of the existence of black holes \cite{LIGOScientific:2016aoc,LIGOScientific:2016sjg}. Subsequently, the Event Horizon Telescope (EHT) collaboration successively released the observational images of the supermassive black hole M87* \cite{EventHorizonTelescope:2019dse} and the Sgr A* \cite{EventHorizonTelescope:2022wkp} at the center of the Milky Way, greatly advancing theoretical research on black holes and ushering in a golden age of black hole physics. Black holes are not only ideal celestial objects for studying the properties of strong gravitational fields and testing gravitational theories, but they also play a vital role in explaining high-energy astrophysical phenomena, such as active galactic nuclei (AGN) \cite{McKinney:2004ka}, gamma-ray bursts (GRB)\cite{Lee:1999se,Tchekhovskoy:2008gq,Komissarov:2009dn} , and ultraluminous X-ray binaries \cite{King:2001pm}. These high-energy astrophysical activities release enormous amounts of energy, so exploring energy extraction from rotating black holes helps reveal the essential properties of high-energy astrophysical phenomena. Christodoulou has for the first time proven that a Kerr black hole with mass $M$ and spin parameter $a$ possesses an irreducible mass \cite{Christodoulou:1970wf}
\begin{equation}
	\begin{aligned}
		M_{irr}=M\sqrt{\frac{1}{2}(1+\sqrt{1-\frac{a^2}{M^2}})},
		\label{eqirr}
	\end{aligned}
\end{equation}
and the corresponding horizon area of this black hole is $A_H=16\pi M_{irr}^2$. Based on the Bekenstein-Hawking black hole entropy area relationship, the horizon area is proportional to the black hole entropy $S=A/4=4\pi M_{irr}^2$ \cite{Bekenstein:1972tm,Hawking:1974rv,Bekenstein:1973ur}. Therefore, the value of irreducible mass is strictly constrained by the second law of black hole thermodynamics, which also sets an upper limit of the rotational energy that a rotating black hole can release outward. For a static black hole with $a=0$, the rotational energy is $E_{rot}=0$. For an extreme Kerr black hole with $a=1$, the maximum rotational energy that can be extracted from the black hole is approximately $E_{rot}\approx0.29M$ \cite{Wei:2022jbi}.

Penrose pioneered the use of thought experiments to propose a mechanism for extracting energy from rotating black holes, known as the Penrose process, the key to which is the conservation law \cite{Penrose:1969pc}. The particle splits into two particles in the ergosphere region; one particle falls into the black hole, while the other escapes to infinity. If the angular momentum of the infalling particle is opposite to the black hole's rotation direction, according to conservation laws, an observer at infinity will see that the infalling particle carries negative energy while the particle escaping to infinity has positive energy, so energy is extracted from the black hole. However, there are two formidable problems associated with the Penrose process. Firstly, the two particles must have a relative three-velocity of $c/2$ after splitting \cite{Bardeen:1972fi}. For this reason, the Penrose process is difficult to realize in realistic astrophysical scenarios. Secondly, the Penrose process lacks an initiation mechanism and has low energy extraction efficiency, making it difficult to explain high energy astrophysical phenomena \cite{Wald:1974kya} . Subsequently, various mechanisms for extracting energy from rotating black holes have been extensively studied, such as the magnetic Penrose process \cite{Wagh:1985vuj}, the Blandford–Znajek (BZ) process \cite{Blandford:1977ds}, the collisional Penrose process \cite{piran1975high}, the magnetohydrodynamic Penrose process \cite{takahashi1990}, Superradiant scattering \cite{Teukolsky:1974yv}. It is worth noting that the BZ process extracts energy from a rotating black hole by utilizing the interaction between the black hole and the surrounding electromagnetic field, and is regarded as the leading mechanism powering relativistic jets of active galactic nuclei \cite{McKinney:2004ka,Hawley:2005xs,Komissarov:2007rc} and gamma-ray bursts \cite{Lee:1999se,Komissarov:2009dn,Tchekhovskoy:2008gq}.

Recently, the EHT collaboration released polarized emission from the supermassive black hole M87*, confirming the existence of a magnetic field around the black hole \cite{EventHorizonTelescope:2021bee,EventHorizonTelescope:2025vum} . This provides observational evidence for extracting the rotational energy of black holes via magnetic reconnection. Magnetic reconnection is an efficient physical process for releasing magnetic energy, which can explain solar and stellar flares \cite{parker1963solar,zimovets2021quasi}, the coronal mass ejections \cite{lin2000effects}, the hot spots generated in jet flows \cite{Aimar:2023kzj,Ripperda:2020bpz} and so on. In fact, Comisso and Asenjo first adopted a purely toroidal magnetic field to discuss the conditions for extracting rotational energy from Kerr black holes via magnetic reconnection \cite{Koide:2008xr}. During magnetic reconnection, the oppositely directed magnetic field lines in the plasma break and reconnect, and magnetic energy can be converted into the kinetic energy and acceleration of particles. Furthermore, the general-relativistic kinetic simulations of black hole magnetospheres indicate that accelerated plasma particles can enter a negative energy orbit from the perspective of an infinite observer \cite{Parfrey:2018dnc}, and particles carrying negative energy will eventually fall into the black hole, thereby extracting the rotational energy of the black hole. In particular, Comisso and Asenjo first investigated the fast magnetic reconnection process in the ergosphere of rotating Kerr black holes and found that, under appropriate conditions, the power of energy extracted via the fast magnetic reconnection process may be higher than that of the BZ process \cite{Comisso:2020ykg}. Subsequently, the extraction of energy from black holes via fast magnetic reconnection has been extensively investigated, such as the rotating charged black holes in Kalb-Ramond gravity \cite{Yao:2026cvs}, the rotating black hole in Bumblebee gravity \cite{YuChih:2025hsg}, the Kerr–Newman black hole in perfect fluid dark matter \cite{Rodriguez:2024jzw}, the Kerr-de Sitter black holes \cite{Wang:2022qmg}, the Kerr-Bertotti-Robinson black hole \cite{Zeng:2025olq} , the Konoplya–Zhidenko rotating non-Kerr black hole \cite{Long:2024tws}, Kerr-Sen-AdS black hole \cite{Zeng:2025vjt} and so on \cite{Zhang:2024rvk,Fan:2024rsa,Fan:2024fcy,Zhao:2025uuk,Kuang:2022ojj}. Currently, researches on extracting rotational energy from black holes through magnetic reconnection processes mainly investigate circular accretion flow matter outside innermost stable circular orbit (ISCO). In fact, no stable circular orbits exist for accretion material within the ISCO, and the plasma will plunge toward the event horizon due to the strong gravitational effect of the black hole \cite{Mummery:2022ana,Mummery:2023tgh,Mummery:2024mrq}. When magnetic reconnection occurs near the event horizon, the energy extraction process of black holes and the dynamical behavior of plasma may provide more information about spacetime backgrounds, offering a new perspective for exploring gravitational theories and testing gravitational properties \cite{Wilkins:2020pgu,Dong:2023bbd} . Therefore, magnetic reconnection in the plunging region has also been investigated for various black hole spacetimes, such as Kerr black hole \cite{Shen:2024sdr}, Kerr–Taub–NUT black hole \cite{Cheng:2025qlc}, Kerr–Bertotti–Robinson black hole \cite{Zeng:2025olq} and so on \cite{Chen:2024ggq,Zeng:2025vjt}. The above results indicate that energy extraction in the plunging region via magnetic reconnection can be realized for black holes with relatively low spin,  and its energy extraction efficiency is higher than that in the circular orbit region.

On the other hand, ModMax electrodynamics is a nonlinear generalization of Maxwell's theory, which introduces an adjustable nonlinear parameter $\gamma$ while maintaining conformal invariance and electromagnetic duality symmetry \cite{Cirilo-Lombardo:2023poc}. ModMax theory can be widely applied to holography \cite{Kosyakov:2020wxv}, high-energy physics \cite{Kuzenko:2024zra} and cosmology \cite{Lechner:2022qhb}. Moreover, ModMax theory has attracted much attention as a highly competitive candidate theoretical model for the photon coupling effect in strong electromagnetic fields \cite{Kosyakov:2020wxv}. In particular, within the framework of the coupling between general relativity and nonlinear electrodynamics, ModMax theory provides an exact solution for a class of static spherically symmetric black holes \cite{Pantig:2022gih,Flores-Alfonso:2020euz}, and using the Newman–Janis algorithm \cite{Newman:1965tw}, the solution for rotating ModMax black holes can be obtained \cite{Karshiboev:2024xxx}. Subsequently, black holes in ModMax theory have been extensively investigated. For example, Ahmad investigated the shadows and quasinormal modes of ModMax black holes and found that the screening factor parameter $\gamma$ increases the shadow radius \cite{Al-Badawi:2025coy}. Karshiboev and Atamurotov investigated the shadows of rotating charged ModMax black hole and found that the screening factor $\gamma$ increases the shadow radius of the black hole and reduces the shadow deformation, while the effect of the charge parameter $Q$ is opposite to that of the parameter gamma $\gamma$ \cite{Karshiboev:2024xxx}. As well as the thermodynamics and optical aspects of ModMax black holes \cite{Al-Badawi:2026ael} , the tidal effects in the vicinity of rotating ModMax black hole \cite{Asgher:2025jxq}, the lensing and quasinormal modes of dysonic ModMax black holes \cite{Pantig:2022gih} and so on \cite{Kurbanov:2026zjb}. The above results indicate that nonlinear electromagnetic modifications can modify the spacetime background of black holes, thereby affecting the gravitational properties in the vicinity of black holes. Therefore, this paper aims to investigate the magnetic reconnection process of rotating ModMax black holes in the circular orbit and plunging regions, analyze the effect of nonlinear electromagnetic modifications on the black hole magnetic reconnection process, and further discuss the influence of parameters $\gamma$ and charge $Q$ on the efficiency and power of energy extraction from black holes.

The remainder of this paper is organized as follows. In Sect. \ref{secspacetime}, we briefly introduce rotating ModMax black hole spacetime. In Sect. \ref{seccircular}, we investigate energy extraction from rotating ModMax black holes in circular orbital regions. In Sect. \ref{secplunging}, we analyze energy extraction from rotating ModMax black holes in the plunging region. Finally, we conclude in Sect. \ref{secconclusion}. Throughout this paper, we will adopt the natural unit with $G = c =1$, and the mass of black hole $M=1$ for simplicity.

\section{Introduction to Rotating ModMax Black Hole Spacetime}\label{secspacetime}

Using the Newman-Janis algorithm \cite{Newman:1965tw}, the metric of a rotating ModMax black hole in Boyer–Lindquist coordinates is given as \cite{Karshiboev:2024zyn,Karshiboev:2024xxx}
\begin{equation}
	ds^2
	= -\frac{\Delta}{\rho^2}
	\left( dt - a \sin^2\theta \, d\phi \right)^2
	+ \frac{\rho^2}{\Delta} dr^2
	+ \rho^2 d\theta^2
	+ \frac{\sin^2\theta}{\rho^2}
	\left( a dt - (r^2+a^2)d\phi \right)^2 .
\end{equation}
where
\begin{equation}
	\rho^2 = r^2 + a^2 \cos^2\theta, ~~~\Delta = r^2 + a^2 - 2Mr + Q^2 e^{-\gamma}.
\end{equation}
Here, $M$, $a$, $Q$, and $\gamma$ represent the mass, rotation parameter, charge, and screening factor of the black hole, respectively. When $\gamma=0$, the solution reduces to the classical Kerr-Newman black hole, while for $\gamma=Q=0$, it recovers the well-known Kerr black hole. The horizon radii of the black hole can be obtained by solving $g^{rr}=0$ as
\begin{equation}
r_{\pm}=M\pm\sqrt{M^2-a^2-Q^2e^{-\gamma}} .
\end{equation}
Furthermore, by solving $g_{tt}=0$, the radii of the inner and outer infinite redshift surfaces of the black hole can be obtained 
\begin{equation}
	r_{e\pm} = M \pm \sqrt{M^2 - a^2\cos^2\theta - Q^2 e^{-\gamma}} .
\end{equation}
Obviously, at the poles $(\theta=0)$, the event horizon coincides with the outer ergosphere, $r_{e+}=r_+=M+\sqrt{M^2-a^2-Q^2e^{-\gamma}}$; at the equator ($\theta=\pi/2$), however, the outer ergosphere lies outside the event horizon, $r_+<r_{e+}= M + \sqrt{M^2 - Q^2 e^{-\gamma}}$. In addition, the magnetic reconnection process is considered to occur on the equatorial plane current sheet located within the ergosphere. Therefore, to ensure that this process can effectively extract the rotational energy of the black hole, the condition $Q^2\le e^\gamma$ must be satisfied.

We now turn to the geodesic equations around the rotating ModMax black hole, which can be derived from the Lagrangian for a massive particle
\begin{equation}
\mathcal{L} = \frac{1}{2} \left( g_{tt} \dot{t}^2 + 2g_{t\phi} \dot{t} \dot{\phi} + g_{rr} \dot{r}^2 + g_{\theta\theta} \dot{\theta}^2 + g_{\phi\phi} \dot{\phi}^2 \right),
\end{equation}
where the dot denotes the derivative with respect to the affine parameter $\tau$. From the time translation invariance and rotational symmetry about the axis of symmetry, two conserved quantities can be obtained: energy $E$ and axial component of the angular momentum $L_z$.
\begin{equation}
	p_t=\frac{\partial \mathcal{L}}{\partial \dot{t}}
	=g_{tt}\dot{t}+g_{t\phi}\dot{\phi}
	=-E,
	\label{eqpt}
\end{equation}
\begin{equation}
	p_\phi=\frac{\partial \mathcal{L}}{\partial \dot{\phi}}
	=g_{\phi\phi}\dot{\phi}+g_{t\phi}\dot{t}
	=L_z.
	\label{eqpphi}
\end{equation}
On the other hand, referring to the model proposed by Comisso and Asenjo \cite{Koide:2008xr}, magnetic reconnection is assumed to occur within the equatorial current sheet. To this end, we focus on the bulk plasma motion in the equatorial plane ($\theta=\pi/2$), which leads to $\dot{\theta}=0$, and take $\mathcal{L}=-\frac{1}{2} \varepsilon$. Here, $\varepsilon=1$ corresponds to massive particles and $\varepsilon=0$ corresponds to  photons. Substituting this condition into Eqs. (\ref{eqpt}) and (\ref{eqpphi}), the first-order differential equations for geodesic motion in the equatorial plane can be obtained
\begin{equation}
	\dot{r}
	=
	\frac{
		\sqrt{
			\left(aL-(a^2+r^2)E\right)^2
			-\Delta\left(\varepsilon^2 r^2+(L-aE)^2\right)
		}
	}{r^2}=R(r),
\label{eqRrr}
\end{equation}
\begin{equation}
	\dot{\phi}
	=
	\frac{1}{r^2}
	\left[
	\frac{a}{\Delta}
	\left(
	E(a^2+r^2)-aL
	\right)
	+
	(L-aE)
	\right],
	\label{eq-phi}
\end{equation}
\begin{equation}
	\dot{t}
	=
	\frac{1}{r^2}
	\left[
	\frac{a^2+r^2}{\Delta}
	\left(
	E(a^2+r^2)-aL
	\right)
	+
	a(L-aE)
	\right].
	\label{eq-tt}
\end{equation}

\begin{figure}[H]
	\centering
	\begin{minipage}{0.25\textwidth}
		\centering
		\includegraphics[width=\linewidth]{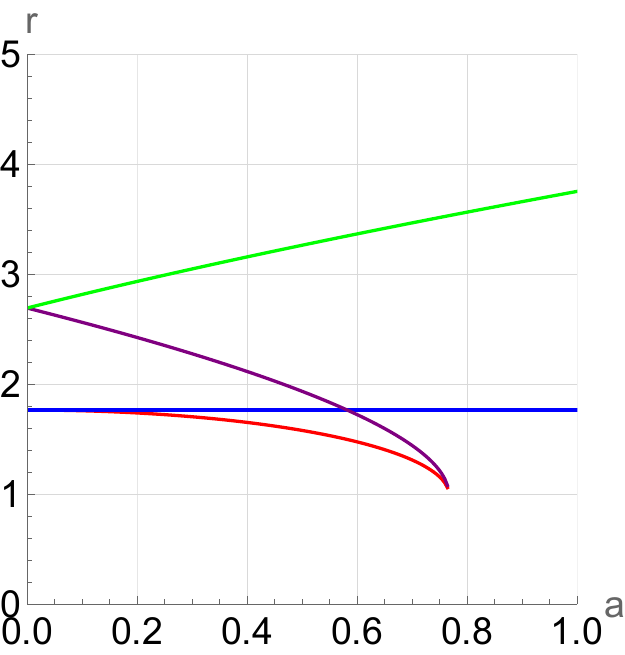}
		{\small(a)$Q=0.5,\ \gamma=-0.5$}
	\end{minipage}
	\begin{minipage}{0.25\textwidth}
		\centering
		\includegraphics[width=\linewidth]{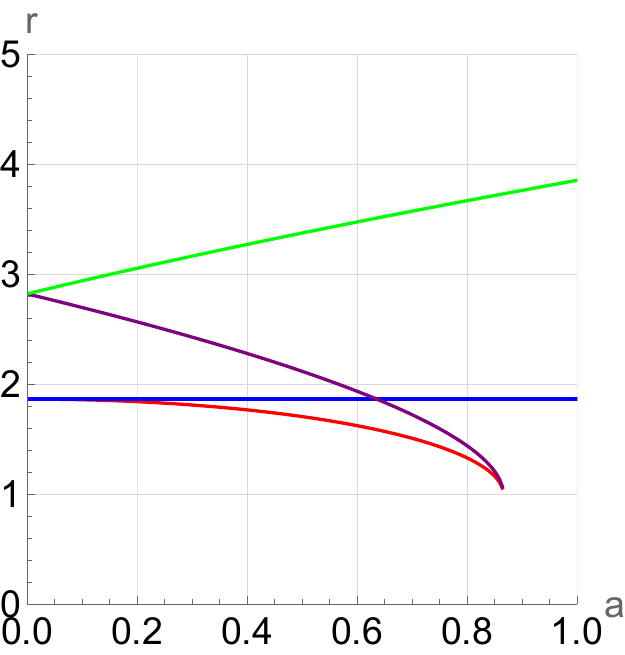}
		{\small(b)$Q=0.5,\ \gamma=0$}
	\end{minipage}
	\begin{minipage}{0.25\textwidth}
		\centering
		\includegraphics[width=\linewidth]{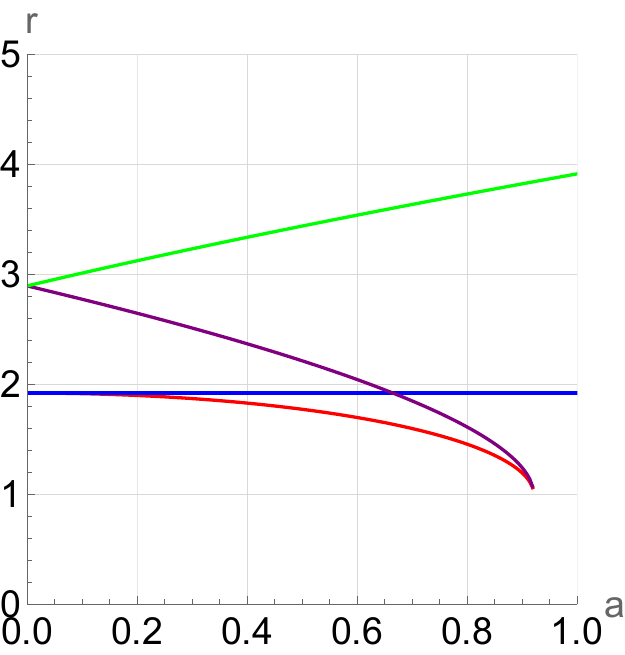}
		{\small(c)$Q=0.5,\ \gamma=0.5$}
	\end{minipage}
	\vspace{0.3cm}
	\begin{minipage}{0.25\textwidth}
		\centering
		\includegraphics[width=\linewidth]{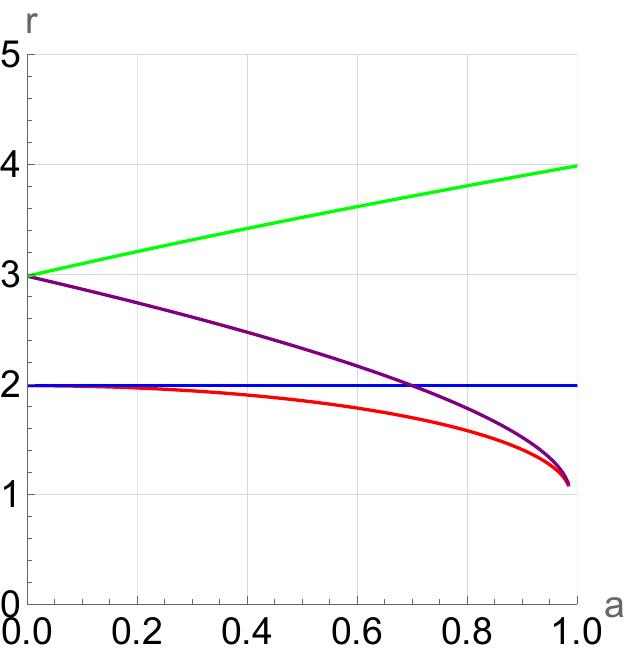}
		{\small(d)$Q=0.2,\ \gamma=0.5$}
	\end{minipage}
	\begin{minipage}{0.25\textwidth}
		\centering
		\includegraphics[width=\linewidth]{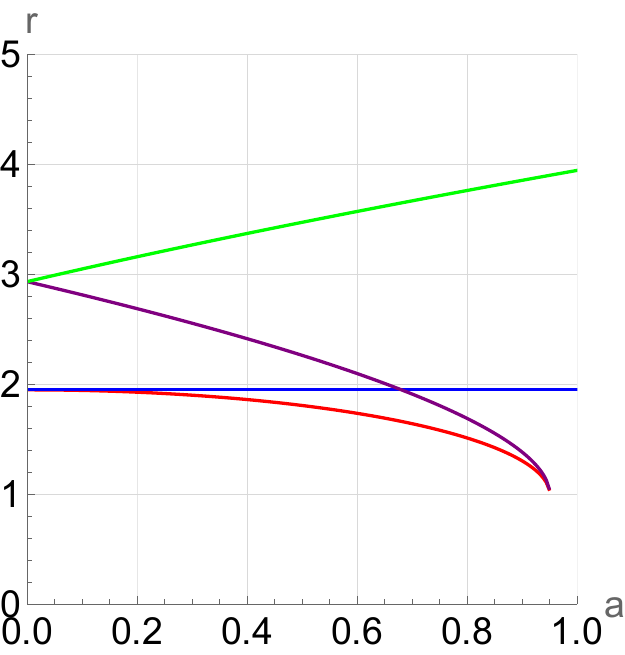}
		{\small(e)$Q=0.4,\ \gamma=0.5$}
	\end{minipage}
	\begin{minipage}{0.25\textwidth}
		\centering
		\includegraphics[width=\linewidth]{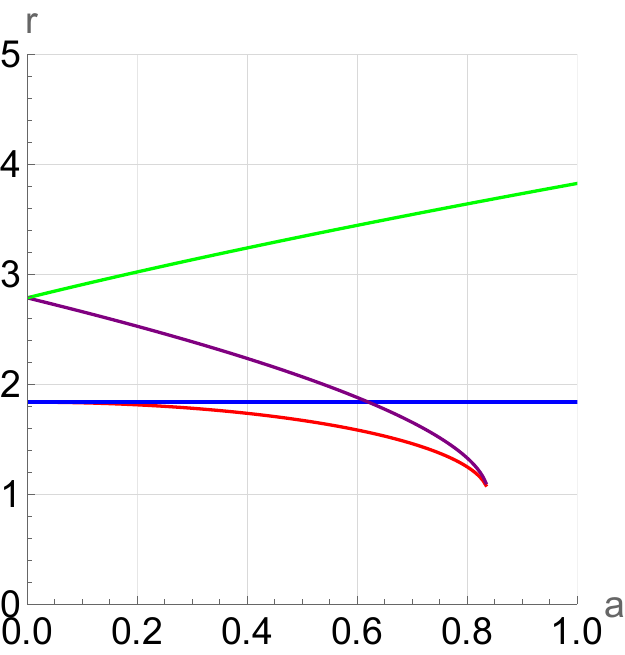}
	{\small(f)$Q=0.7,\ \gamma=0.5$}
	\end{minipage}
	
	\caption{The radii of the prograde and retrograde circular photon orbits, photon sphere radius, boundary of ergosphere $r_{e+}$, and event horizon $r_+$ as functions of spin parameter $a$. Here, the green, purple, blue, and red curves represent the retrograde circular photon orbit, prograde circular photon orbit, ergosphere boundary $r_{e+}$, and event horizon $r_+$, respectively.}
	\label{fig:radius-spin}
\end{figure}

Next, we focus on circular photon orbits in the equatorial plane. For photons, $\varepsilon=0$, and circular orbits require both the radial velocity and radial acceleration to vanish
\begin{equation}
	R(r)=0, ~~~~ \qquad R'(r)=0.
	\label{eqRr}
\end{equation}
where a prime denotes a derivative with respect to $r$. The radii of the prograde and retrograde circular photon orbits in the equatorial plane are directly obtained from Eq. (\ref{eqRr}). Since the expressions for the radii of circular photon orbits are relatively complex, we present in Fig. \ref{fig:radius-spin} the prograde and retrograde circular photon orbit radii in the equatorial plane as functions of the spin parameter $a$. Here, the green, purple, blue, and red curves represent the radii of retrograde circular photon orbit, the prograde circular photon orbit, the ergosphere boundary, and the event horizon, respectively. Clearly, as the spin parameter $a$ increases, the radius of the event horizon gradually decreases, the prograde circular photon orbit radius decreases, while the retrograde circular photon orbit radius increases. As $a$ increases further, the prograde circular photon orbit radius continuously approaches the event horizon and eventually coincides with it in the extremal ModMax black hole case. It is worth noting that both a larger $\gamma$ and a smaller $Q$ increase the maximum allowed spin parameter $a$, exhibiting a competitive effect between the two parameters. Conversely, a smaller $\gamma$ and a larger $Q$ significantly reduce the maximum spin parameter value, allowing the black hole to reach the extremal state at a lower spin parameter $a$, which is more feasible for energy extraction under low spin conditions. Since the energy extraction process occurs inside the ergosphere, and according to Fig. \ref{fig:radius-spin}, only photons on prograde circular orbits can enter the ergosphere, we therefore only need to focus on prograde orbits. Furthermore, at the circular photon orbit radius, massive particles cannot maintain circular orbital motion with finite energy, so the effective energy extraction actually occurs between the prograde circular photon orbit radius and the ergosphere boundary. For massive particles, the Keplerian angular velocity of the prograde circular orbit is given by \cite{Cai:2023ygh}.

\begin{equation}
	\Omega =\frac{d\phi/d\tau}{dt/d\tau}=
	\frac{
		-\partial_r g_{t\phi}
		+
		\sqrt{
			\left(\partial_r g_{t\phi}\right)^2
			-
			\left(\partial_r g_{tt}\right)
			\left(\partial_r g_{\phi\phi}\right)
		}
	}{
		\partial_r g_{\phi\phi}
	} .
\end{equation}
Clearly, the Keplerian angular velocity of the particle involves only the azimuthal component of the angular velocity, and it can be obtained directly from Eqs. (\ref{eq-phi}) and (\ref{eq-tt}).

\section{Extracting Energy from Rotating ModMax Black Holes in Circular Orbital Regions}\label{seccircular}

\subsection{Magnetic Reconnection Process in Circular Orbits}\label{sec:circular1}

The magnetic reconnection process in the equatorial plane follows the Comisso-Asenjo framework \cite{Comisso:2020ykg}. In this framework, when the current sheet in the equatorial plane exceeds a critical aspect ratio \cite{Uzdensky:2014uda,Comisso:2017arh}, magnetic flux ropes or plasmoids are formed \cite{Daughton:2009tr,Bhattacharjee:2009fp}, which subsequently trigger fast magnetic reconnection. Numerical simulations show that there always exists a dominant X-point during magnetic reconnection, at which the magnetic field lines reconnect \cite{Parfrey:2018dnc,Ripperda:2020bpz}. In this process, the magnetic energy is converted into plasma energy, causing the decelerated particles,  which carry negative energy, to fall into the black hole, while the accelerated particles, which carry positive energy, escape from the ergosphere to infinity, thus extracting energy from the rotating black hole. Next, we will briefly review the specific contents of the Comisso-Asenjo framework.

\begin{figure}[H]
	\centering
	\includegraphics[
	width=0.6\textwidth
	]{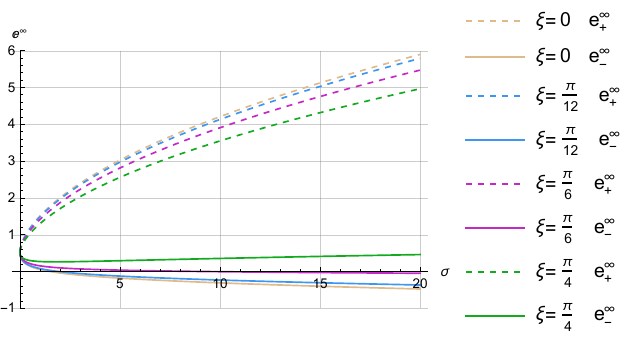}
	\hspace{0.02\textwidth}
	\caption{The behaviors of $e_{+}^{\infty}$ and $e_{-}^{\infty}$ with the magnetization parameter $\sigma$ for different azimuthal angle $\xi$. Here, we fix $a=0.9$, $Q=0.5$, $\gamma=-0.5$, and $r=1.5$}
	\label{fig-energytwo}
\end{figure}

\begin{figure}[H]
	\centering
	\begin{minipage}{0.48\textwidth}
		\centering
		\includegraphics[width=\linewidth]{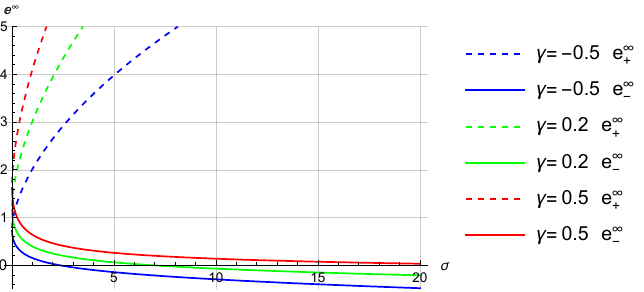}
		
		{\small (a)$a=0.9,\ Q=0.3,\ r=1.5,\ \xi=\frac{\pi}{12}$}
	\end{minipage}
	\hfill
	\begin{minipage}{0.48\textwidth}
		\centering
		\includegraphics[width=\linewidth]{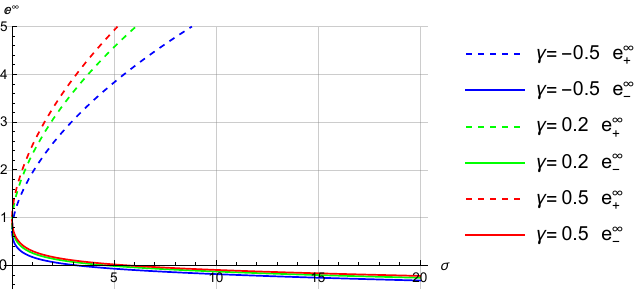}
		{\small (b)$a=0.9,\ Q=0.3,\ r=1.6,\ \xi=\frac{\pi}{12}$}
	\end{minipage}
	\caption{The behaviors of $e_{+}^{\infty}$ and $e_{-}^{\infty}$ with the magnetization parameter $\sigma$ for different screening factors $\gamma$.}
	\label{figtwo-panels}
\end{figure}

\begin{figure}[H]
	\centering
	\includegraphics[width=0.6\textwidth]{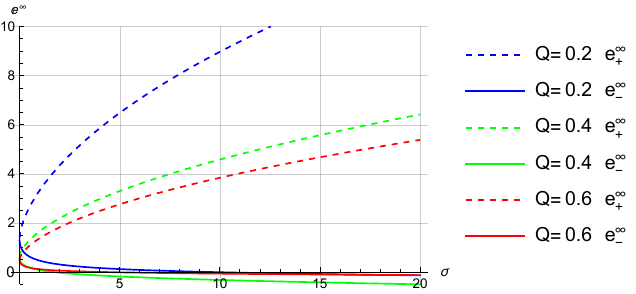}
\caption{The behaviors of $e_{+}^{\infty}$ and $e_{-}^{\infty}$ with the magnetization parameter $\sigma$ for different charge parameters $Q$. Here, we fix $a=0.9$, $\gamma=-0.5$, $r=1.5$, and $\xi=\frac{\pi}{12}$.}
	\label{figsingle}
\end{figure}

Here, it is necessary to investigate the energy density of the plasma. For convenience, we adopt the zero-angular-momentum-observer (ZAMO) frame \cite{Bardeen:1972fi}. In this frame, the metric can be rewritten as
\begin{equation}
	ds^2 = -d\hat{t}^{\,2} + \sum_{i=1}^{3} \left(d\hat{x}^{\,i}\right)^2
	= \eta_{\alpha\beta} d\hat{x}^{\alpha} d\hat{x}^{\beta},
\end{equation}
where the coordinate transformations are given by
\begin{equation}
	d\hat{t} = \alpha dt, 
	\qquad
	d\hat{x}^{\,i} = \sqrt{g_{ii}}\,dx^i - \alpha \beta^i dt ,
\end{equation}
here, $\alpha$, $\beta$, and $\omega^\phi$ are the lapse function, shift vector components, and angular velocity, respectively, and they are given as
\begin{equation}
	\beta^\phi = \frac{\sqrt{g_{\phi\phi}}\,\omega^\phi}{\alpha},
	\qquad
	\alpha = \left( -g_{tt} + \frac{g_{t\phi}^2}{g_{\phi\phi}} \right)^{1/2},
	\qquad
	\omega^\phi = -\frac{g_{t\phi}}{g_{\phi\phi}},\qquad
	\beta^r=\beta^\theta=0.
\end{equation}
It is worth noting that in the ZAMO frame, the Kepler velocity of particle is given by
\begin{equation}
	\hat{v}_{K} =
	\frac{\sqrt{g_{\phi\phi}}\,\Omega - \alpha\beta^\phi}{\alpha}.
\end{equation}
Following reference \cite{Zeng:2025olq}, we adopt the single-fluid plasma approximation, where the energy-momentum tensor is defined as
\begin{equation}
	T^{\mu\nu}
	= p g^{\mu\nu}
	+ w u^\mu u^\nu
	+ F^{\mu}_{ \sigma}F^{\nu\sigma}
	-\frac{1}{4}g^{\mu\nu}F^{\alpha\beta}F_{\alpha\beta}.
\end{equation}
where $p$, $u^\mu$, $w$ and $F_{\alpha\beta}$ are the pressure, four-velocity, plasma enthalpy density and  electromagnetic field tensor, respectively. In the Comisso-Asenjo framework, the magnetic reconnection process occurring in the equatorial plane of a rotating black hole is considered highly efficient, meaning that all magnetic energy is converted into the kinetic energy of the plasma, while the contribution from the electromagnetic field tensor can be neglected. In addition, by further introducing the adiabatic and incompressible approximations of the plasma, the ratio of the energy density to the enthalpy density at infinity for the accelerated and decelerated plasma streams can be expressed as \cite{Comisso:2020ykg}
\begin{equation}
	e_{\pm}^{\infty}
	= \alpha \hat{\gamma}_{K}
	\left[
	\left(1+\beta^\phi \hat{v}_{K}\right)(1+\sigma)^{1/2}
	\pm \cos\xi \left(\beta^\phi+\hat{v}_{K}\right)\sigma^{1/2}
	-\frac{1}{4}
	\frac{
		(1+\sigma)^{1/2}
		\mp \cos\xi \hat{v}_{K}\sigma^{1/2}
	}{
		\hat{\gamma}_{K}^{2}
		\left(1+\sigma-\cos^2\xi\,\hat{v}_{K}^{2}\sigma\right)
	}
	\right].
	\label{eqeint}
\end{equation}
Here, $\xi$ denotes the azimuthal angle of the plasma in the local rest frame, $\sigma=B^2/w$ is the plasma magnetization parameter, and $\hat{\gamma}_{K}$ is the Lorentz factor of $\hat{v}_{K}$, which can be written as
\begin{equation}
	\hat{\gamma}_{K} = \left(1-\hat{v}_{K}^{2}\right)^{-1/2}.
\end{equation}
In an extremely relativistic hot plasma, whose thermodynamic equation of state is given by $w=4p$, the process of extracting energy from a rotating black hole via magnetic reconnection is similar to the Penrose particle-splitting mechanism, and the plasma system is required to satisfy two conditions \cite{Penrose:1969pc}
\begin{equation}
	e_{-}^{\infty}<0,
	\qquad
	\Delta e_{+}^{\infty}
	=
	e_{+}^{\infty}
	-
	\left[
	1-\frac{\Gamma}{4(\Gamma-1)}
	\right]
	=
	e_{+}^{\infty}>0,
\end{equation}
where, $\Gamma$ is the polytropic index, similar to reference \cite{Zeng:2025vjt}, and we fix $\Gamma=4/3$. Furthermore, we show $e_{+}^{\infty}$ and $e_{-}^{\infty}$ in the Figs. \ref{fig-energytwo}, \ref{figtwo-panels} and \ref{figsingle}, where $r$ denotes the dominant X-point of magnetic reconnection.

As shown in Fig. \ref{fig-energytwo}, $e_{+}^{\infty}$ is always greater than $0$, while $e_{-}^{\infty}$ may be either positive or negative. Therefore, whether energy can be extracted from the rotating ModMax black hole through the magnetic reconnection process depends on the sign of $e_{-}^{\infty}$. The sign of $e_{-}^{\infty}$ is closely related to both the magnetization parameter $\sigma$ and the azimuthal angle $\xi$. As the magnetization parameter $\sigma$ increases, $e_{-}^{\infty}$ gradually decreases, while a larger azimuthal angle $\xi$ leads to a larger value of $e_{-}^{\infty}$. This implies that a smaller azimuthal angle is more favorable for energy extraction, a conclusion that is consistent with findings for other types of black holes \cite{Wei:2022jbi,Zeng:2025olq,Long:2024tws,Zeng:2025vjt}.

In Fig. \ref{figtwo-panels}, for a fixed magnetization parameter $\sigma$, $e_{-}^{\infty}$ increases gradually with increasing $\gamma$. Moreover, the variation of $e_{-}^{\infty}$ with $\gamma$ becomes more obvious for smaller magnetic reconnection point $r$. In addition, in Fig. \ref{figsingle}, there is no monotonic relationship between  $e_{\pm}^{\infty}$ and charge parameter Q, which is closely related to the location of the reconnection point.

\subsection{Parameter Space for Energy Extraction in Circular Orbits}\label{sec:circular2}

\begin{figure}[H]
	\centering
	
	\begin{minipage}{0.32\textwidth}
		\centering
		\includegraphics[width=\linewidth]{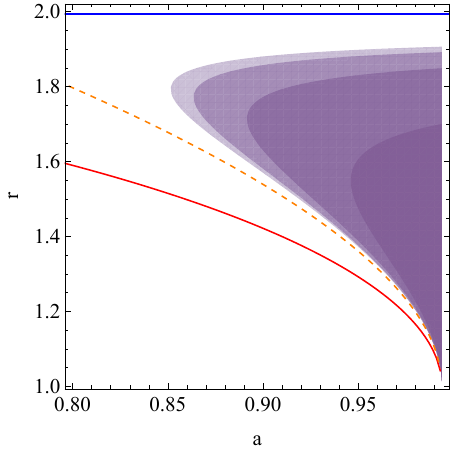}
		
		\vspace{0.1cm}
		{\small (a) $Q=0.1,\ \gamma=-0.2,\ \xi=\frac{\pi}{12}$}
	\end{minipage}
	\hfill
	\begin{minipage}{0.32\textwidth}
		\centering
		\includegraphics[width=\linewidth]{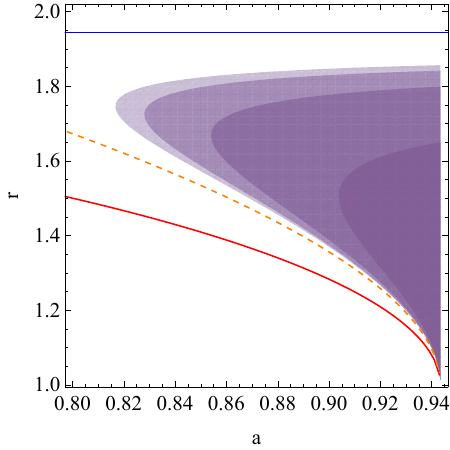}
		
		\vspace{0.1cm}
		{\small (b) $Q=0.3,\ \gamma=-0.2,\ \xi=\frac{\pi}{12}$}
	\end{minipage}
	\hfill
	\begin{minipage}{0.32\textwidth}
		\centering
		\includegraphics[width=\linewidth]{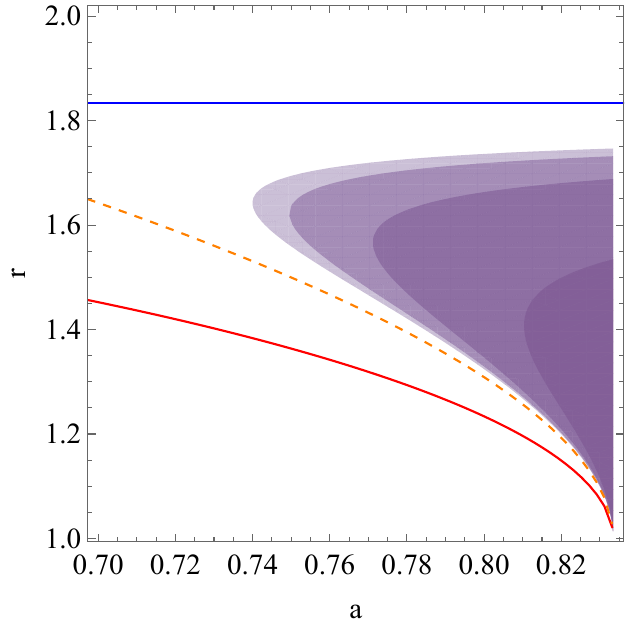}
		
		\vspace{0.1cm}
		{\small (c) $Q=0.5,\ \gamma=-0.2,\ \xi=\frac{\pi}{12}$}
	\end{minipage}
	
	\caption{The allowed regions for energy extraction in the parameter space $(a,r)$ for different charge parameters $Q$.}
	\label{fig:parameter-space0}
\end{figure}

\begin{figure}[H]
	\centering
	\begin{minipage}{0.32\textwidth}
		\centering
		\includegraphics[width=\linewidth]{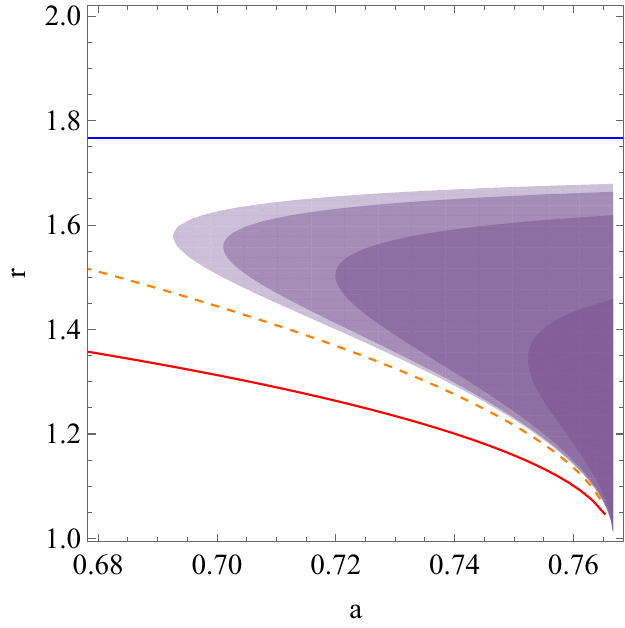}
		\vspace{0.1cm}
		{\small (a) $Q=0.5,\ \gamma=-0.5,\ \xi=\frac{\pi}{12}$}
	\end{minipage}
	\hfill
	\begin{minipage}{0.32\textwidth}
		\centering
		\includegraphics[width=\linewidth]{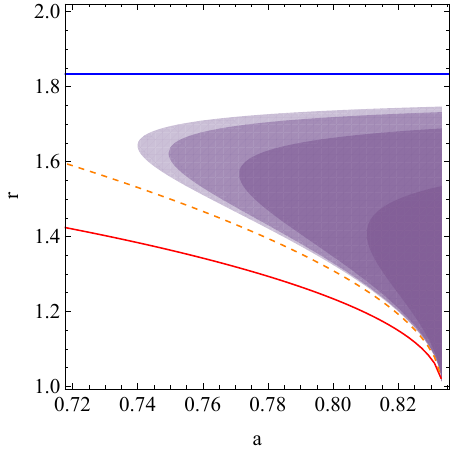}
		\vspace{0.1cm}
		{\small (b) $Q=0.5,\ \gamma=-0.2,\ \xi=\frac{\pi}{12}$}
	\end{minipage}
	\hfill
	\begin{minipage}{0.32\textwidth}
		\centering
		\includegraphics[width=\linewidth]{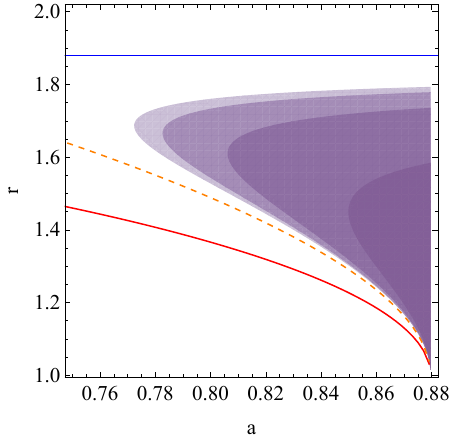}
		\vspace{0.1cm}
		{\small (c) $Q=0.5,\ \gamma=0.1,\ \xi=\frac{\pi}{12}$}
	\end{minipage}
	\caption{The allowed regions for energy extraction in the parameter space $(a,r)$ for different screening factors $\gamma$.}
	\label{fig:parameter-space1}
\end{figure}

\begin{figure}[H]
	\centering
	\begin{minipage}{0.32\textwidth}
		\centering
	\includegraphics[width=\linewidth]{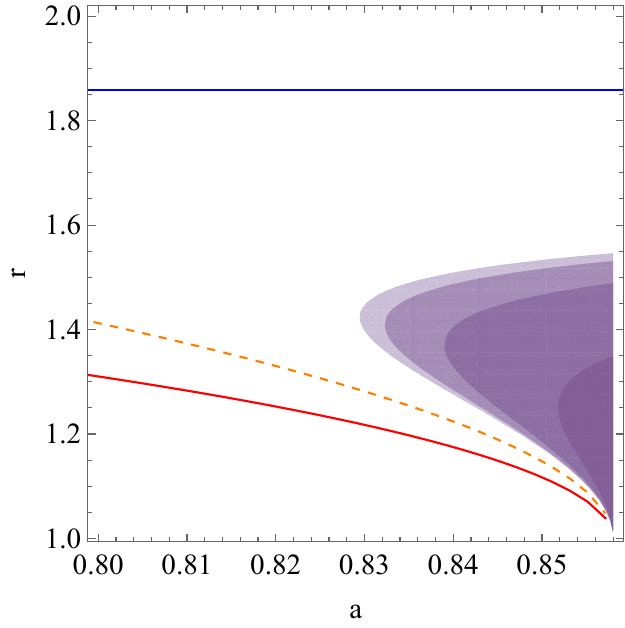}
		\vspace{0.1cm}
		{\small (a) $Q=0.4,\ \gamma=-0.5,\ \xi=\frac{\pi}{6}$}
	\end{minipage}
	\hfill
	\begin{minipage}{0.32\textwidth}
		\centering
	\includegraphics[width=\linewidth]{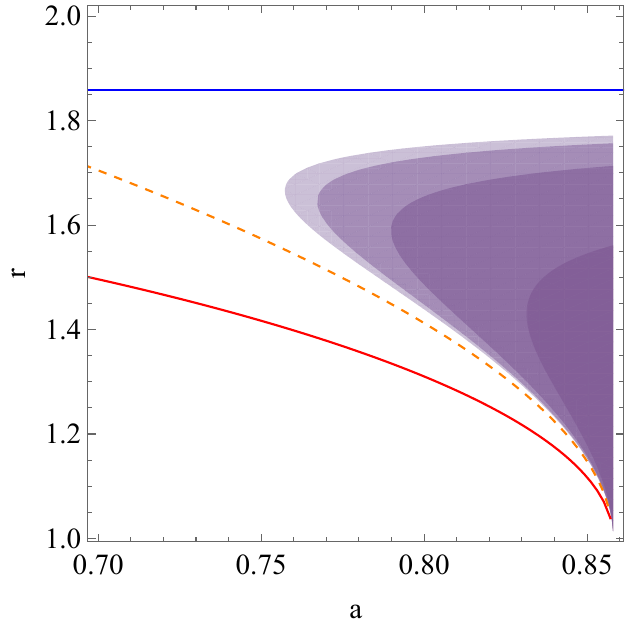}
		\vspace{0.1cm}
		{\small (b) $Q=0.4,\ \gamma=-0.5,\ \xi=\frac{\pi}{12}$}
	\end{minipage}
	\hfill
	\begin{minipage}{0.32\textwidth}
		\centering
	\includegraphics[width=\linewidth]{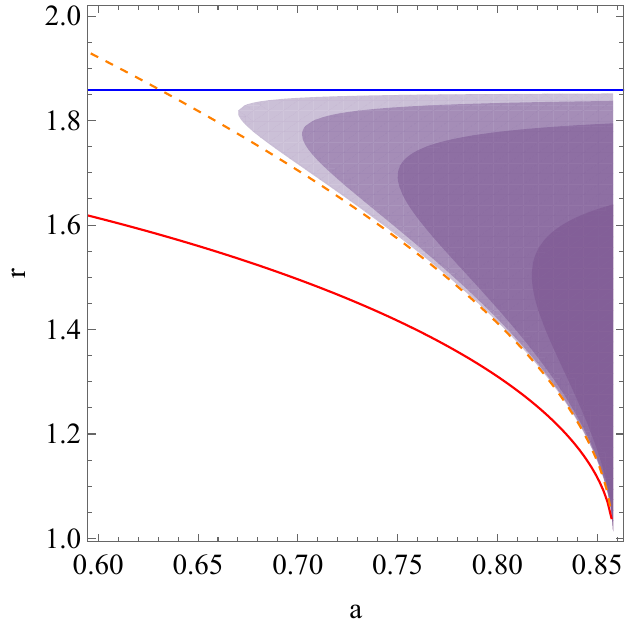}
		\vspace{0.1cm}
		{\small (c) $Q=0.4,\ \gamma=-0.5,\ \xi=0$}
	\end{minipage}
	\caption{The allowed regions for energy extraction in the parameter space $(a,r)$ for different azimuthal angles $\xi$.}
	\label{fig:parameter-space2}
\end{figure}

\begin{figure}[H]
	\centering
\includegraphics[width=0.5\textwidth]{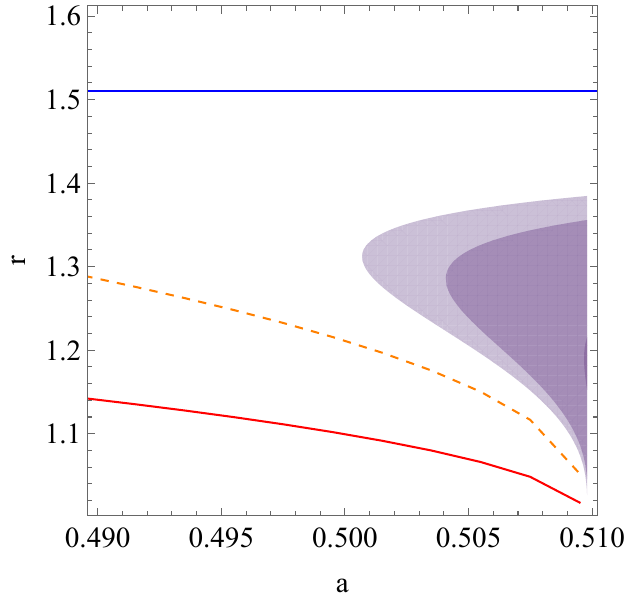}
\caption{The allowed regions for energy extraction in the parameter space $(a,r)$ for $Q=0.67$, $\gamma=-0.5$, and $\xi=\pi/12$. From left to right, the shaded regions correspond to $\sigma=100$ and $30$, respectively.}
	\label{fig:single3}
\end{figure}

Next, we investigate the allowed energy extraction region of the rotating ModMax black hole in the 
$(a,r)$ parameter space, which requires $e_{-}^{\infty}<0$. In Figs. \ref{fig:parameter-space0}-\ref{fig:single3}, the red solid curve, red dashed curve, and blue solid curve represent the event horizon, the radius of the circular photon orbit, and the ergosphere boundary $r_{e+}$, respectively. In addition, the shaded regions from left to right correspond to $\sigma=100$, $30$, $10$, and $3$, respectively, and the energy extraction region is always located between the circular photon orbit and the ergosphere boundary $r_{e+}$.

In Fig. \ref{fig:parameter-space0}, as the plasma magnetization parameter $\sigma$ decreases, the allowable region of energy extraction also shrinks. Furthermore, for larger charge parameters $Q$, the energy extraction region shrinks and shifts toward smaller $r$. It is worth noting that as the charge parameter $Q$ increases, both the maximum and minimum spin parameters $a$ allowed for energy extraction decrease. For example, from Fig. (\ref{fig:parameter-space0}a) to Fig. (\ref{fig:parameter-space0}c), the maximum allowed spin parameter decreases from $a\simeq0.9938$ to $a\simeq0.8334$, while the minimum allowed spin parameter also decreases from $a\simeq0.8517$ to $a\simeq0.7401$. Therefore, the charge parameter $Q$ lowers the threshold of the spin parameter $a$ required for energy extraction from the rotating ModMax black hole, thus allowing energy extraction to be achieved at smaller values of $a$.

In Fig. \ref{fig:parameter-space1}, as screening factor $\gamma$ increases, the energy extraction region becomes larger and shifts toward larger $r$. In particular, when screening factor $\gamma$ decreases, both the maximum and minimum spin parameters $a$ allowed for energy extraction decrease, an effect similar to that of increasing  the charge parameter $Q$. Therefore, both a smaller $\gamma$ and a larger $Q$ can lower the threshold of the spin parameter $a$ required for energy extraction from a rotating ModMax black hole.

In Fig. \ref{fig:parameter-space2}, we investigate the allowed regions for energy extraction in the parameter space $(a,r)$ for different azimuthal angles $\xi$. It can be seen that as $\xi$ decreases, the region for energy extraction gradually expands and moves in the direction of increasing $r$, and the minimum spin parameter $a$ that allows energy extraction also decreases as $\xi$ decreases. However, we are more concerned with the effects of $\gamma$ and $Q$ on energy extraction, so in the following analysis, we fix $\xi=\pi/12$ without loss of generality.

As discussed above, both a smaller $\gamma$ and a larger $Q$ reduce the minimum spin parameter $a$ required for energy extraction. To investigate the lower bound of $a$, we set $\gamma=-0.5$ and $Q=0.67$ in Fig. \ref{fig:single3}, a parameter combination that satisfies the black hole existence condition $a^2+Q^2e^{-\gamma}\leq 1$. The results show that the minimum spin parameter that allows energy extraction is $a\simeq0.50071$, indicating that it is feasible to extract energy from a circular orbit via magnetic reconnection at moderate spins.

\subsection{Power and Efficiency of Energy Extraction in Circular Orbits}\label{sec:circular3}

In this section, we investigate the power and efficiency of energy extraction from the rotating ModMax black hole. The power of energy extraction is given by \cite{Comisso:2020ykg}
\begin{equation}
	P=-e_{-}^{\infty} w A_{\rm in} U_{\rm in},
\end{equation}
where $w$ is the enthalpy density of the plasma, and $U_{\rm in}$ is the plasma inflow velocity. It should be emphasized that $U_{\rm in}\approx 0.1$ for the collisionless case \cite{Comisso:2016}, while $U_{\rm in}\approx 0.01$ for the collisional case \cite{Huang:2010fj,Uzdensky:2010}. Here, we consider the latter collisionless case. $A_{\rm in}$ is the cross-sectional area of the inflowing plasma, which can be given by
\begin{equation}
	A_{\rm in}\sim \left(r_{e+}^2-r_{\rm ph}^2\right),
\end{equation}
here $r_{e+}$ is the outer boundary of the ergosphere and $r_{\rm ph}$ is the radius of the circular photon orbit in the equatorial plane.

\begin{figure}[H]
	\centering
	\subfigure[\, $Q=0.4,\gamma=-0.5,\xi=\pi/12,a=0.85$]
	{\includegraphics[width=7cm]{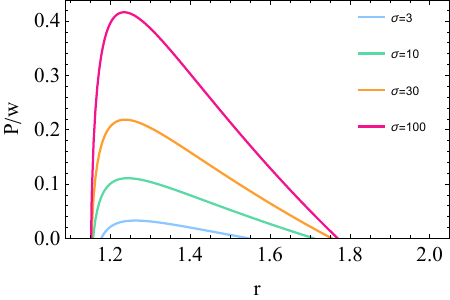}} \hspace{2mm}
	\subfigure[\, $\sigma=100,\gamma=-0.2,\xi=\pi/12,a=0.85$]
	{\includegraphics[width=7cm]{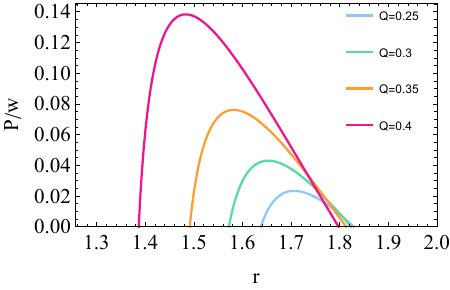}}\\
	\subfigure[\, $\sigma=100,Q=0.4,\xi=\pi/12,a=0.85$]
	{\includegraphics[width=7cm]{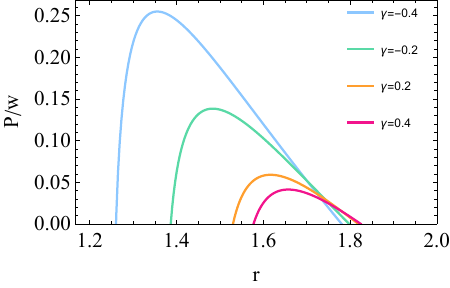}}\hspace{2mm}
	\subfigure[\, $\sigma=100,Q=0.4,\gamma=-0.4,\xi=\pi/12$]
	{\includegraphics[width=7cm]{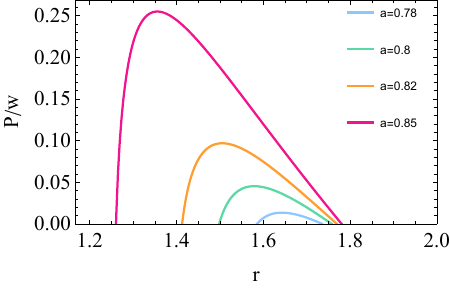}}\\
	\caption{The energy extraction power per unit enthalpy density $P/w$ as a function of the dominant X-point location $r$.}
	\label{figpower}
\end{figure}

In Fig. \ref{figpower}, we present the energy extraction power per enthalpy density as a function of the the dominant X-point location $r$. Clearly, the energy extraction power is always located between the circular photon orbit and the ergosphere boundary $r_{e+}$, which is consistent with the previous discussion. As the magnetic reconnection position $r$ increases, the power first increases and then decreases. For a larger magnetization parameter $\sigma$, the power of energy extraction is also larger. Furthermore, as the screening factor $\gamma$ increases, the peak of the energy extraction power decreases, while a larger charge parameter $Q$ gives a higher peak power, indicating that there is a competitive effect between $\gamma$ and $Q$. However, as the $r$ increases further, the energy extraction power corresponding to a larger $Q$ and a smaller $\gamma$ is lower. This is because the magnetic reconnection position $r$ for energy extraction decreases with increasing $Q$ and decreasing $\gamma$. In addition, as the spin parameter $a$ increases, the power of energy extraction decreases.

\begin{figure}[H]
	\centering
	\subfigure[\, $Q=0.4,\gamma=-0.5,\xi=\pi/12,a=0.85$]
	{\includegraphics[width=7cm]{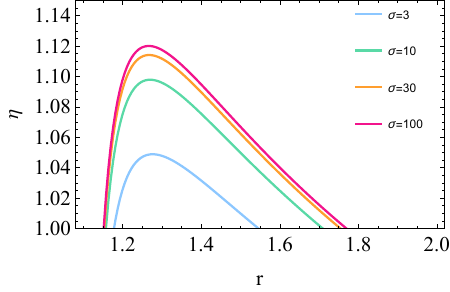}} \hspace{2mm}
	\subfigure[\,$\sigma=100,\gamma=-0.2,\xi=\pi/12,a=0.85$]
	{\includegraphics[width=7cm]{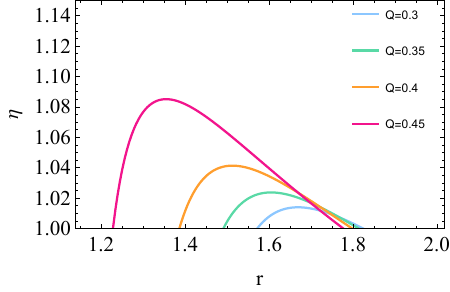}}\\
	\subfigure[\, $\sigma=100,Q=0.4,\xi=\pi/12,a=0.85$]
	{\includegraphics[width=7cm]{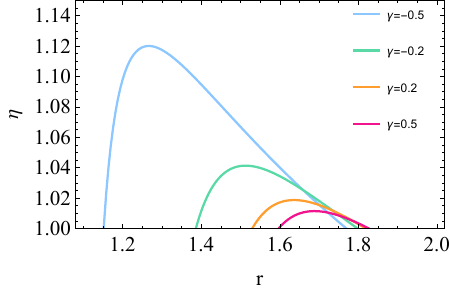}}\hspace{2mm}
	\subfigure[\, $\sigma=100,Q=0.4,\gamma=-0.5,\xi=\pi/12$]
	{\includegraphics[width=7cm]{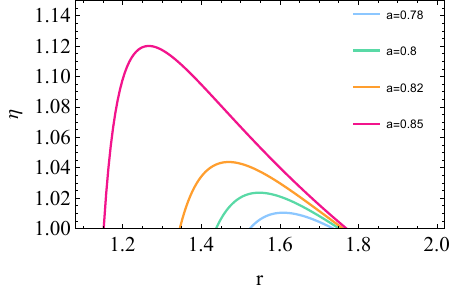}}\\
	\caption{The energy extraction efficiency $\eta$ as a function of the dominant X-point location $r$.}
	\label{figefficiency}
\end{figure}

Next, we investigate the efficiency of energy extraction from the rotating ModMax black hole, which is given by \cite{Comisso:2020ykg}
\begin{equation}
	\eta = \frac{e_{+}^{\infty}}{e_{+}^{\infty}+e_{-}^{\infty}}.
\end{equation}
Obviously, since $e_{-}^{\infty}<0$ and $e_{+}^{\infty}>0$, the efficiency of energy extraction is always greater than 1. In Fig. \ref{figefficiency}, we present the energy extraction efficiency $\eta$ as a function of the dominant X-point location $r$. As the dominant X-point location $r$ increases, the energy extraction efficiency first increases and then decreases. The efficiency of energy extraction increases with increasing magnetization parameter $\sigma$. The dependence of the efficiency on $Q$ and $\gamma$ is similar to that of the power: a larger $Q$ and a smaller $\gamma$ lead to a higher peak efficiency. Furthermore, the energy extraction efficiency also becomes higher with increasing $a$.

\section{Extracting Energy from Rotating ModMax Black Holes in the Plunging Region}\label{secplunging}

\subsection{Magnetic Reconnection Process in the Plunging Region}\label{sec:plunging1}

As discussed above, outside the ISCO ($r>r_I$, here $r_I$ is the ISCO radius), the plasma moves on circular orbits in the equatorial plane of the black hole. Once the plasma crosses the ISCO, it quickly plunges into the black hole, since no stable circular orbits exist inside the ISCO. In the plunging region $(r_+<r<r_I)$, the energy extraction efficiency and power of the black hole differ significantly from those in the circular orbit case \cite{Zeng:2025vjt,Zeng:2025olq,Shen:2024sdr}. Therefore, this section focuses on investigating the energy extraction process of plasma in the plunging region. It is worth noting that, unlike the circular orbit case, the plasma velocity in the plunging region has a radial component, and thus Eq. (\ref{eqeint}) needs to be modified. Here, we adopt the ZAMO frame, and the relation between the four-velocity in the ZAMO frame and that in the Boyer-Lindquist coordinates can be given as
\begin{equation}
	\hat{U}^{\mu}
	=
	\hat{\gamma}_{s}
	\left\{1,\hat{v}_{s}^{(r)},0,\hat{v}_{s}^{(\phi)}\right\}
	=
	\left\{
	\frac{E-\omega^{\phi}L_z}{\alpha},
	\sqrt{g_{rr}}\,U^{r},
	0,
	\frac{L_z}{\sqrt{g_{\phi\phi}}}
	\right\},
	\label{eqUU}
\end{equation}
where the radial component of the four-velocity
\begin{equation}
	 U^{r}=dr/d\tau
\end{equation}
and $dr/d\tau$ is obtained from Eq. (\ref{eqRrr}). The plasma enters the plunging region from the ISCO, with its initial energy and angular momentum given by $E=E({r_I})$ and $L_z=L_z({r_I})$, respectively. Furthermore, $E({r_I})$ and $L({r_I})$ can be obtained using Eq. (\ref{eqRr}) and $R''(r)=0$. Therefore, the four-velocity is rewritten as
\begin{equation}
	(U^{r})^2=(dr/d\tau)^2=-\frac{\left(aL_z(r_I)-(a^2+r^2)E(r_I)\right)^2-\Delta\left(\varepsilon^2 r^2+(L_z(r_I)-aE(r_I))^2\right)}{r^4},
	\label{equur}
\end{equation}
where, the negative sign indicates the inward motion of the plasma. By combining Eqs. (\ref{eqUU}) and (\ref{equur}), $\hat{v}_{s}^{(\phi)}$ and $\hat{v}_{s}^{(r)}$ can be obtained. Furthermore, the energy density per enthalpy is given by \cite{Chen:2024,Chen:2024ggq,Comisso:2020ykg}
\begin{equation}
	e_{\pm}^{\infty}
	=
	\alpha \hat{\gamma}_{s}\gamma_{\rm out}
	\left[
	\left(1+\beta^\phi \hat{v}_{s}^{(\phi)}\right)
	\pm v_{\rm out}
	\left(
	\hat{v}_{s}
	+\beta^\phi
	\frac{\hat{v}_{s}^{(\phi)}}{\hat{v}_{s}}
	\right)\cos\xi
	\mp v_{\rm out}\beta^\phi
	\frac{\hat{v}_{s}^{(r)}}{\hat{\gamma}_{s}\hat{v}_{s}}
	\sin\xi
	\right]
	-
	\frac{\alpha}{4\hat{\gamma}_{s}\gamma_{\rm out}
		\left(1\pm\hat{v}_{s}v_{\rm out}\cos\xi\right)}.
\end{equation}
Here $\hat{v}_s=\sqrt{(\hat{v}_s^{r})^2+(\hat{v}_s^{\phi})^2}$, $\hat{\gamma}_s$ is the Lorentz factor of $\hat{v}_s$. In addition, $v_{out}$ is the speed of outflow and $\gamma_{out}$ is the Lorentz factor of $v_{out}$, which can be defined using the magnetization parameter $\sigma$
\begin{equation}
	v_{out}=\sqrt{\frac{\sigma}{\sigma+1}},~~~\gamma_{out}=\sqrt{1+\sigma}.
\end{equation}

\begin{figure}[H]
	\centering
	\begin{minipage}{0.4\textwidth}
		\centering
		\includegraphics[width=0.9\linewidth]{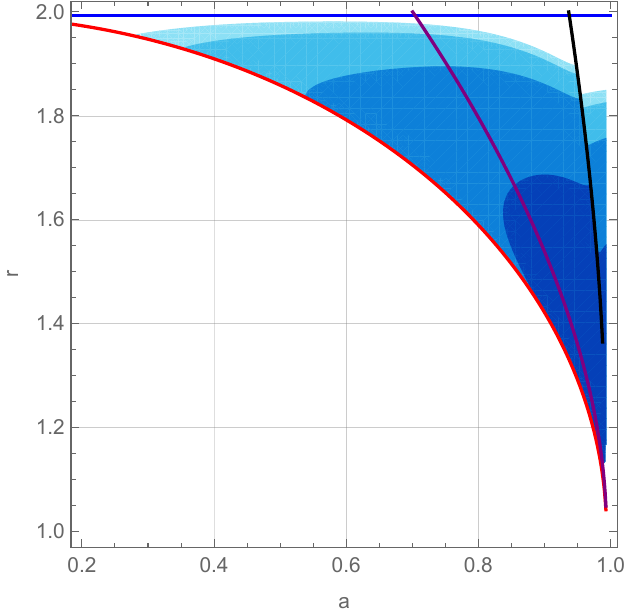}
		\vspace{0.08cm}
		{\small (a)$Q=0.1,\ \gamma=-0.2,\ \xi=\frac{\pi}{12}$}
	\end{minipage}
\hspace{0.03\textwidth}
	\begin{minipage}{0.4\textwidth}
		\centering
		\includegraphics[width=0.9\linewidth]{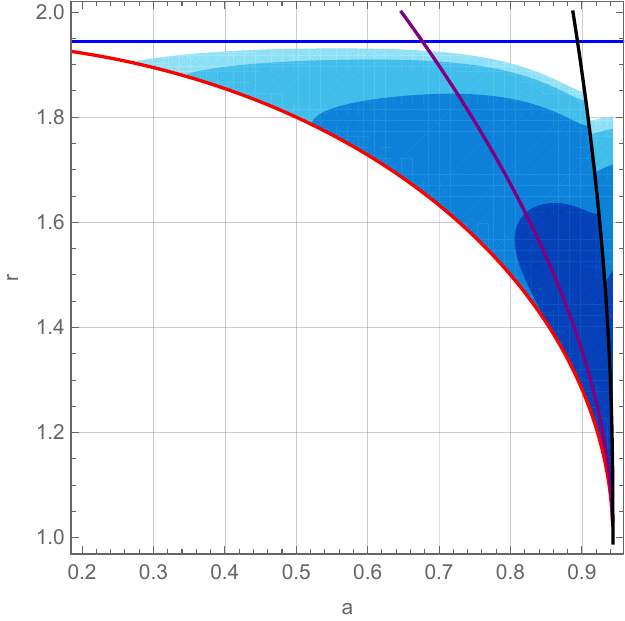}
		\vspace{0.08cm}
		{\small (b)$Q=0.3,\ \gamma=-0.2,\ \xi=\frac{\pi}{12}$}
	\end{minipage}
	\vspace{0.25cm}
	\begin{minipage}{0.4\textwidth}
		\centering
		\includegraphics[width=0.9\linewidth]{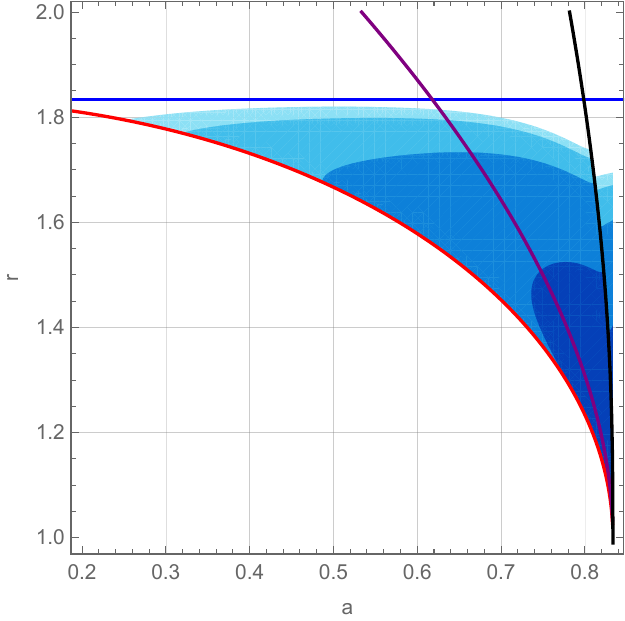}
		\vspace{0.08cm}
		{\small (c)$Q=0.5,\ \gamma=-0.2,\ \xi=\frac{\pi}{12}$}
	\end{minipage}
\hspace{0.03\textwidth}
	\begin{minipage}{0.4\textwidth}
		\centering
		\includegraphics[width=0.9\linewidth]{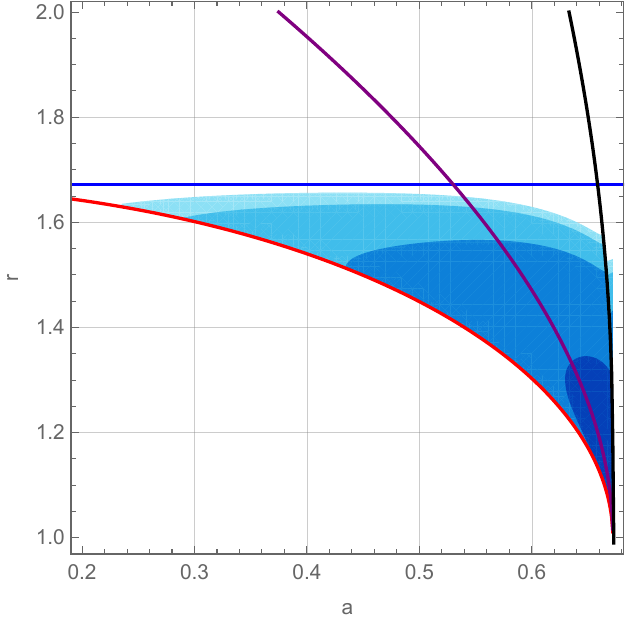}
		\vspace{0.08cm}
		{\small (d)$Q=0.67,\ \gamma=-0.2,\ \xi=\frac{\pi}{12}$}
	\end{minipage}
	\caption{The allowed regions for energy extraction in the parameter space $(a,r)$ of plunging orbits with different charge parameters $Q$.}
	\label{figplungingQ}
\end{figure}

\begin{figure}[H]
	\centering
	\begin{minipage}{0.4\textwidth}
		\centering
		\includegraphics[width=0.9\linewidth]{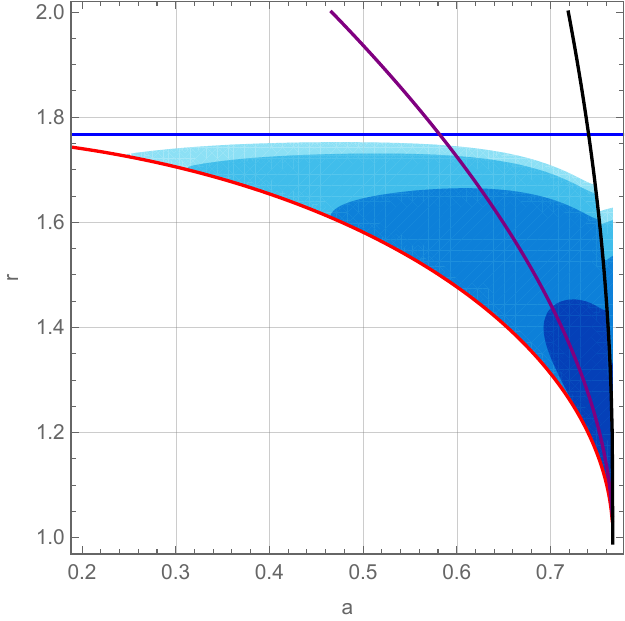}
		\vspace{0.08cm}
		{\small (a)$Q=0.5,\ \gamma=-0.5,\ \xi=\frac{\pi}{12}$}
	\end{minipage}
\hspace{0.015\textwidth}
	\begin{minipage}{0.4\textwidth}
		\centering
		\includegraphics[width=0.9\linewidth]{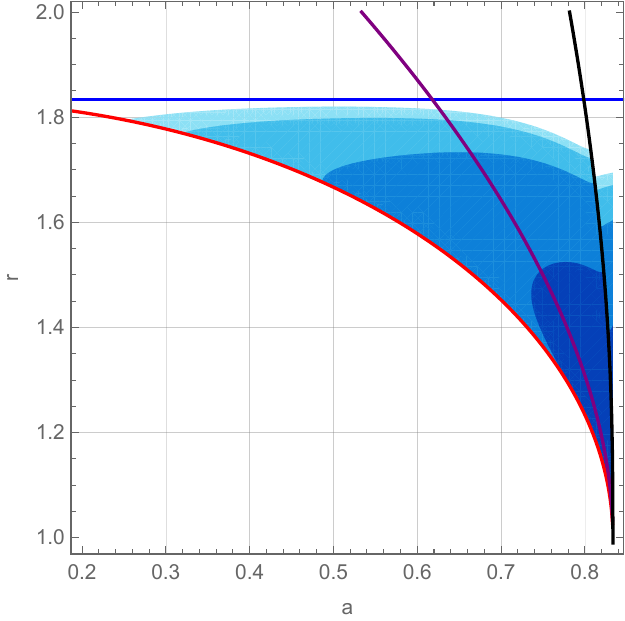}
		\vspace{0.08cm}
		{\small (b)$Q=0.5,\ \gamma=-0.2,\ \xi=\frac{\pi}{12}$}
	\end{minipage}
	\vspace{0.25cm}
	\begin{minipage}{0.4\textwidth}
		\centering
		\includegraphics[width=0.9\linewidth]{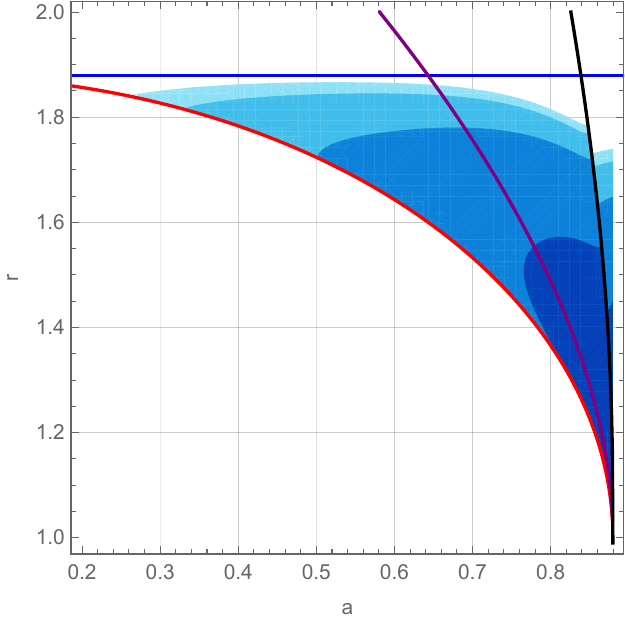}
		\vspace{0.08cm}
		{\small (c)$Q=0.5,\ \gamma=0.1,\ \xi=\frac{\pi}{12}$}
	\end{minipage}
\hspace{0.015\textwidth}
	\begin{minipage}{0.4\textwidth}
		\centering
		\includegraphics[width=0.9\linewidth]{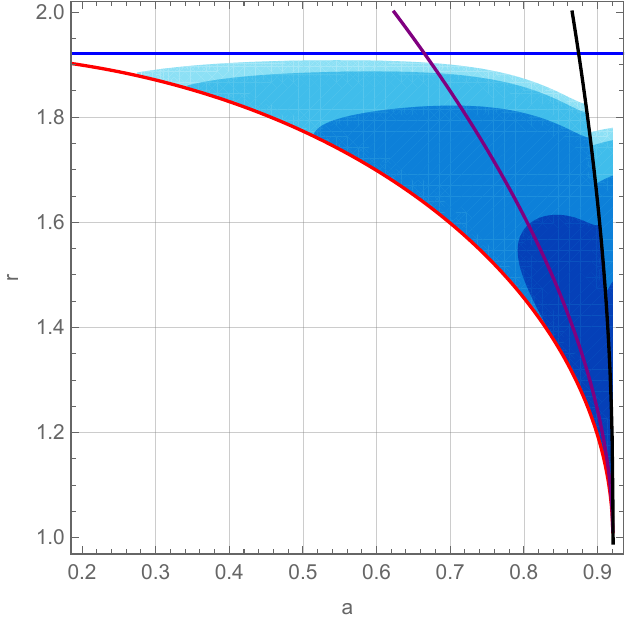}
		\vspace{0.08cm}
		{\small (d)$Q=0.5,\ \gamma=0.5,\ \xi=\frac{\pi}{12}$}
	\end{minipage}
\caption{The allowed regions for energy extraction in the parameter space $(a,r)$ of plunging orbits with different screening factors $\gamma$.}
	\label{figplungingG}
\end{figure}

\begin{figure}[h]
	\centering
	\begin{minipage}{0.32\textwidth}
		\centering
		\includegraphics[width=\linewidth]{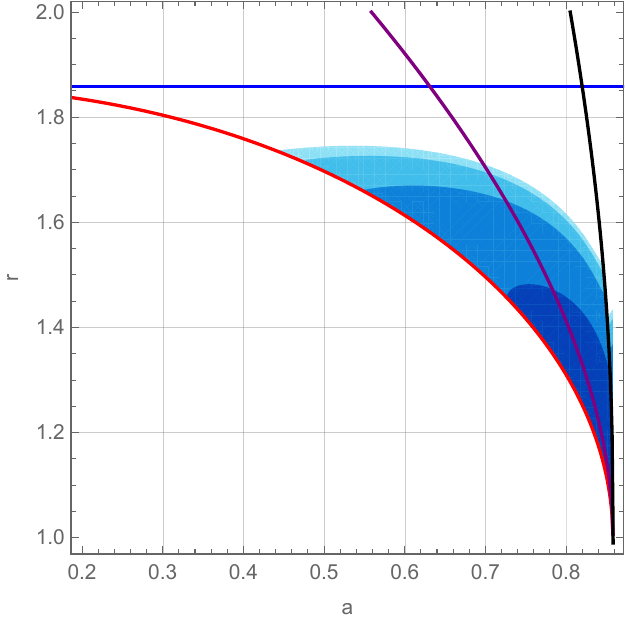}
		\vspace{0.08cm}
		{\small (a)$Q=0.4,\ \gamma=-0.5,\ \xi=\frac{\pi}{6}$}
	\end{minipage}
	\hfill
	\begin{minipage}{0.32\textwidth}
		\centering
		\includegraphics[width=\linewidth]{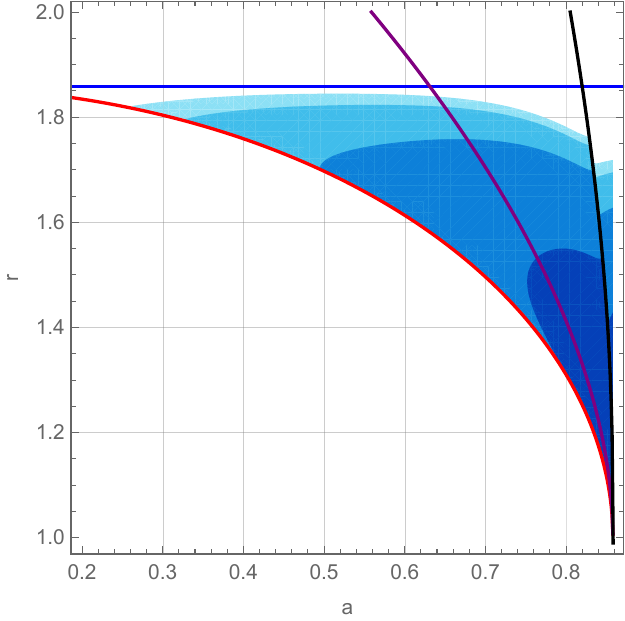}
		\vspace{0.08cm}
		{\small (b)$Q=0.4,\ \gamma=-0.5,\ \xi=\frac{\pi}{12}$}
	\end{minipage}
	\hfill
	\begin{minipage}{0.32\textwidth}
		\centering
		\includegraphics[width=\linewidth]{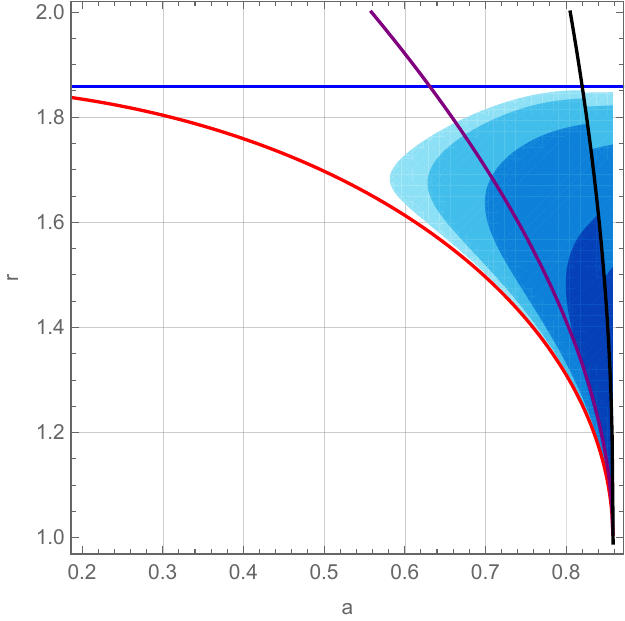}
		\vspace{0.08cm}
		{\small (c)$Q=0.4,\ \gamma=-0.5,\ \xi=0$}
	\end{minipage}
	\caption{The allowed regions for energy extraction in the parameter space $(a,r)$ of plunging orbits with different azimuthal angles $\xi$.}
	\label{figplungingxi}
\end{figure}

\begin{figure}[H]
	\centering
	\includegraphics[width=0.5\textwidth]{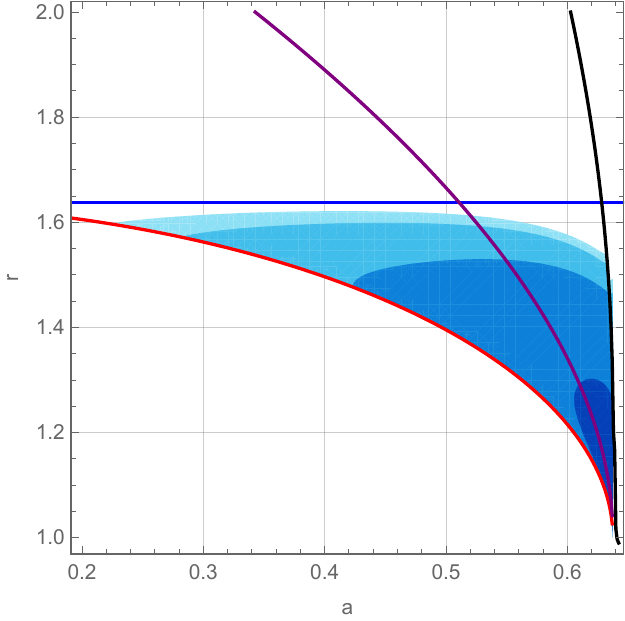}
	\caption{The allowed regions for energy extraction of plunging orbits in the parameter space $(a,r)$ for $Q=0.6$, $\gamma=-0.5$, and $\xi=\pi/12$.}
	\label{figplunginglow}
\end{figure}

Next, the allowed regions for energy extraction of plunging orbits for the rotating ModMax black hole are shown in Figs. \ref{figplungingQ}-\ref{figplunginglow}. The blue, red, purple, and black curves represent the ergosphere boundary $r_{e+}$, the event horizon, the circular photon orbit, and the ISCO, respectively. Furthermore, the shaded regions, ranging from left to right with colors transitioning from light to dark, correspond to the parameter $\sigma=100$, $30$, $10$, and $3$, respectively. A comparison of Figs. \ref{fig:parameter-space0} and \ref{figplungingQ} shows that the allowed region for energy extraction in plunging orbits is significantly larger than that in circular orbits. In the plunging orbit case, the dominant X-point location $r$ for energy extraction can be closer to the ergosphere boundary $r_{e+}$. It is worth noting that for plunging orbits, energy extraction can occur in the region between the circular photon orbit and the event horizon. This is precisely why the minimum spin parameter $a$ allowing for energy extraction in plunging orbits is lower than that for circular orbits. Furthermore, the energy extraction region expands as $\sigma$ increases, whereas it shrinks and shifts toward smaller $r$ values as Q increases. This trend is consistent with that for circular orbits. Furthermore, as the parameter $Q$ increases, the minimum spin parameter allowing for energy extraction decreases from $a\approx0.27839$ in Fig. \ref{figplungingQ}(a) to $a\approx0.22892$ in Fig. \ref{figplungingQ}(d). This indicates that, for plunging orbits, a larger $Q$ also lowers the minimum spin $a$ required for energy extraction from the rotating ModMax black hole.

In Fig. \ref{figplungingG}, as $\gamma$ increases, the energy extraction region expands, which is closely related to the increase in the ergosphere boundary $r_{e+}$. Similar to the case of circular orbits, a smaller parameter $\gamma$ allows for a lower minimum spin parameter for energy extraction. For plunging orbits, the screening factor $\gamma$ and charge parameter $Q$ still exhibit a competing effect on energy extraction. In Fig. \ref{figplungingxi}, we investigate the effect of the azimuthal angle on the energy extraction region. Similar to the case of circular orbits, a smaller azimuthal angle corresponds to a lower minimum spin parameter $a$. However, for $\xi=0$, the energy extraction region shrinks abruptly, a behavior that has been extensively discussed \cite{Zeng:2025vjt,Zeng:2025olq,Shen:2024sdr}. Both smaller values of $\gamma$ and larger values of $Q$ reduce the minimum spin parameter $a$ that allows for energy extraction. Furthermore, in Fig. \ref{figplunginglow}, we investigate the minimum allowed spin parameter in the plunging region. Specifically, fixing $\gamma=-0.5$ and $Q=0.67$, the minimum allowed spin parameter is $a\approx0.22407$. This result indicates that, for plunging orbits, energy extraction from the rotating ModMax black hole is still feasible even at relatively low spins.

\subsection{Power and Efficiency of Energy Extraction in the Plunging Region}\label{sec:plunging2}

In this section, we discuss the efficiency and power of energy extraction in the plunging region. However, for circular orbits, the effects of charge parameter $Q$ and screening factor $\gamma$ on the efficiency and power have already been analyzed in detail. Therefore, this section does not further investigate the effects of $\gamma$ and $Q$ in the plunging region. We are more interested in the difference in energy extraction efficiency and power between the circular and plunging regions. The expressions for the energy extraction efficiency and power in the plunging region are the same as those for circular orbits, so they are not repeated here. However, in the plunging region, the plasma plunges from the ISCO into the black hole, and its cross-sectional area differs from that in the circular orbit case, which should be modified as follows
\begin{equation}
	A_{\rm in}\sim \left(r_{e+}^2-r_{+}^2\right).
\end{equation}

\begin{figure}[H]
	\centering
\includegraphics[width=0.5\textwidth]{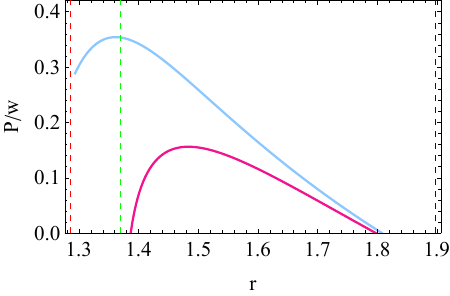}
	\caption{The energy extraction power per unit enthalpy density $P/w$ as a function of the dominant X-point location $r$. The blue and red curves represent the energy extraction power for plunging and circular orbits, respectively. The red, green, and black dashed curves correspond to the event horizon, the circular photon orbit, and the ergosphere boundary $r_{e+}$, respectively. Here we fix $Q=0.4$, $\gamma=-0.2$, $\xi=\pi/12$, $\sigma=100$, and $a=0.85$.}
	\label{figpow}
\end{figure}

\begin{figure}[H]
	\centering	\includegraphics[width=0.5\textwidth]{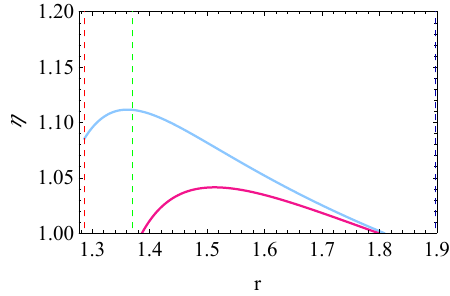}
	\caption{The energy extraction efficiency $\eta$ as a function of the dominant X-point location $r$. The blue and red curves represent the energy extraction power for plunging and circular orbits, respectively. The red, green, and black dashed curves correspond to the event horizon, the circular photon orbit, and the ergosphere boundary $r_{e+}$, respectively. Here we fix $Q=0.4$, $\gamma=-0.2$, $\xi=\pi/12$, $\sigma=100$, and $a=0.85$.}
	\label{figeff}
\end{figure}

Subsequently, we show the energy extraction efficiency and power in the plunging region in Figs. \ref{figpow} and \ref{figeff}, respectively, and compare them with those for circular orbits. It is found that the efficiency and power of energy extraction in the plunging region are similar to those of circular orbits, initially increasing and then decreasing as $r$ increases. It is worth noting that both the efficiency and power of energy extraction in the plunging region are higher than those for circular orbits, and this difference is more pronounced near the circular photon orbit.

\section{Conclusion}\label{secconclusion}

Polarimetric observations of the supermassive black hole M87* have confirmed the presence of magnetic fields around the black hole \cite{EventHorizonTelescope:2021bee,EventHorizonTelescope:2025vum}, providing important observational evidence for energy extraction from rotating black holes via magnetic reconnection. As a candidate theory describing photon interactions in strong electromagnetic fields, ModMax theory has attracted widespread attention in fields such as holography, high-energy physics, and cosmology \cite{Cirilo-Lombardo:2023poc,Kosyakov:2020wxv,Lechner:2022qhb}. In this paper, we investigate the magnetic reconnection process around the rotating ModMax black hole within the framework of ModMax theory.

First, we investigate the event horizon, the radius of the ergosphere ($r_{e+}$), and the retrograde and prograde circular photon orbits. As the spin parameter $a$ increases, the radii of the event horizon and the prograde circular photon orbit decrease, while the radius of the retrograde circular photon orbit increases. For the extremal ModMax black hole, the prograde circular photon orbit coincides with the event horizon. In particular, increasing the screening factor $\gamma$ or decreasing the charge parameter $Q$ raises the maximum allowed spin parameter $a$. Conversely, smaller $\gamma$ or larger $Q$ significantly reduce the maximum allowed spin parameter, causing ModMax black holes to approach the extremal limit at lower spins, thereby facilitating energy extraction at lower spins. Subsequently, we investigate the ratio of the energy density to enthalpy density at infinity for the accelerated and decelerated plasma streams, and examine the effects of the parameters $\xi$, $\sigma$, $\gamma$, and $Q$ on energy extraction. The results indicate that a larger magnetization parameter $\sigma$ or a smaller azimuthal angle $\xi$ is more favorable for energy extraction. Furthermore, a smaller $\gamma$ or a larger $Q$ also facilitates energy extraction from a rotating ModMax black hole, which is consistent with the above conclusions.

Furthermore, we analyzed the energy extraction mechanisms for ModMax black holes in the circular orbit and plunging regions. For the allowed parameter space $(a,r)$ for energy extraction, the effects of the parameters $\sigma$, $\gamma$, $Q$, and $\xi$ on the circular and plunging orbit regions are consistent. A larger $\sigma$ and a smaller $\xi$ increase the energy extraction region. However, a smaller $\gamma$ and a larger $Q$ reduce the energy extraction region, and the parameters $Q$ and $\gamma$ exhibit a competing effect on the energy extraction region. It is worth noting that a larger $Q$ and a smaller $\gamma$ lower the spin parameter $a$ required for energy extraction. Therefore, for specific parameter values, energy extraction via magnetic reconnection is feasible in the ModMax black hole with $a\simeq 0.50071$ in the circular orbit case. In particular, for plunging orbits, the threshold for the spin parameter further decreases to $a\simeq0.22407$, indicating that energy extraction from lower spin black holes is theoretically feasible.

Finally, we investigate the power and efficiency of energy extraction in circular and plunging orbits. A larger $\sigma$ or spin parameter $a$ improves both the power and efficiency of energy extraction. Moreover, a larger $Q$ or a smaller $\gamma$ increases the peaks of the power and efficiency of energy extraction, which is consistent with the previous findings. More importantly, we find that energy extraction region, power and efficiency of energy extraction in plunging orbits are always higher than those in circular orbits. Therefore, energy extraction from a rotating ModMax black hole via magnetic reconnection in plunging orbits may be more favorable.

\begin{acknowledgments}
This work is partly supported by the National Natural Science Foundation of China Grants No. 12505059, China Postdoctoral Science Foundation Grants No. 2025MD784184 and China West Normal University under Grant No. 25KE032.
\end{acknowledgments}

\bibliography{ref}
\bibliographystyle{apsrev}

\end{document}